\documentclass[%
showkeys,
 amsmath,amssymb,
]{revtex4-1}

\usepackage{graphicx}
\usepackage{dcolumn}
\usepackage{bm}

\usepackage{CJK}

\usepackage{color}
\usepackage{amssymb}
\usepackage{amsfonts}

\usepackage[colorlinks,urlcolor=blue,linkcolor=blue,citecolor=blue,anchorcolor=blue]{hyperref}

\begin{document}

\preprint{APS/123-QED}

\title{Modulation of the photonic band structure topology of a honeycomb lattice in an atomic vapor}

\author{Yiqi Zhang$^1$}
\email{zhangyiqi@mail.xjtu.edu.cn}
\author{Xing Liu$^1$}
\author{Milivoj R. Beli\'c$^{2}$}
\email{milivoj.belic@qatar.tamu.edu}
\author{Zhenkun Wu$^1$}
\author{Yanpeng Zhang$^{1}$}
\email{ypzhang@mail.xjtu.edu.cn}
\affiliation{%
 $^1$Key Laboratory for Physical Electronics and Devices of the Ministry of Education \& Shaanxi Key Lab of Information Photonic Technique,
Xi'an Jiaotong University, Xi'an 710049, China \\
$^2$Science Program, Texas A\&M University at Qatar, P.O. Box 23874 Doha, Qatar \\
$^3$Department of Electronic and Information Engineering, Shunde Polytechnic, Shunde 528300, China
}%

\date{\today}

\begin{abstract}
  \noindent
In an atomic vapor, a honeycomb lattice can be constructed by utilizing the three-beam interference method.
  In the method, the interference of the three beams splits the dressed energy level periodically,
  forming a periodic refractive index modulation with the honeycomb profile.
  The energy band topology of the honeycomb lattice can be modulated by frequency detunings,
  thereby affecting the appearance (and disappearance) of Dirac points and cones in the momentum space.
  This effect can be usefully exploited for the generation and manipulation of topological insulators.
\end{abstract}

\maketitle

%

\section{Introduction}

A honeycomb lattice is not a Bravais lattice \cite{peleg.prl.98.103901.2007,ablowitz.pra.79.053830.2009,bahat-treidel.prl.104.063901.2010},
which makes it fundamentally different from the hexagonal lattice \cite{terhalle.prl.101.013903.2008,terhalle.prl.106.083902.2011}.
In recent years, honeycomb lattices in different physical systems \cite{tarruell.nature.483.302.2012,gomes.nature.483.306.2012,rechtsman.nature.496.196.2013,rechtsman.np.7.153.2013,rechtsman.prl.111.103901.2013,plotnik.nm.13.57.2014,jotzu.nature.515.237.2014,song.nc.6.6272.2015} have attracted an increased attention from research community.
An especially interesting development occurred when it was realized that photonic topological insulators can be produced in honeycomb lattices \cite{rechtsman.nature.496.196.2013}.
Floquet topological insulators can confine the propagating light beam at the edges without scattering energy into the bulk,
thus becoming robust against scattering from defects.
Such useful properties result from the topologically protected edge states,
which are mainly determined by the topological structure in the momentum space and are protected by the symmetry.
The edge states are quite stable and independent of the surface structure of the material \cite{kane.prl.95.226801.2005, hsieh.nature.452.970.2008, hasan.rmp.82.3045.2010}.
Recently, unconventional edge states \cite{plotnik.nm.13.57.2014} and pseudospin \cite{song.nc.6.6272.2015} of honeycomb lattices have also been demonstrated.
The novel findings related to honeycomb lattices, as well as photonic topological insulators, can have potential applications
in quantum computing \cite{qi.rmp.83.1057.2011}, optical modulators \cite{liu.nature.474.64.2011} and optical diodes \cite{liang.prl.110.203904.2013}.
Honeycomb lattices can be conveniently formed by the femtosecond laser writing technique or by the optically induced method.
The first method is exclusively used in solid materials \cite{rechtsman.nature.496.196.2013, plotnik.nm.13.57.2014},
while the second method can be used in both solid \cite{song.nc.6.6272.2015} and gaseous \cite{zhang.lpr.9.1.2015} materials.

As far as we know, honeycomb lattices are mostly investigated in solid materials
\cite{singha.science.332.1176.2011,rechtsman.nature.496.196.2013,plotnik.nm.13.57.2014, song.nc.6.6272.2015}.
In addition to solid materials, honeycomb lattices are also reported in gaseous materials
\cite{zhao.prl.97.230404.2006,wu.prb.77.235107.2008,jotzu.nature.515.237.2014, zhang.lpr.9.1.2015}.
However, some extra conditions are required for the hexagonal lattice to form in such materials.
In atomic vapors, when the three-beam interference pattern serves as the dressing field, the dressed atomic states will exhibit controllable optical properties.
In the last decade, periodically dressed atomic vapors -- such as rubidium vapor -- were intensively investigated,
leading to the observation of many interesting effects, including enhanced multi-wave mixing (MWM) signals due to
Bragg reflection of photonic band gap (PBG) structures \cite{artoni.prl.96.073905.2006}.
Also, the Talbot effect of MWM \cite{zhang.ieee.4.2057.2012} and the nonreciprocity of light \cite{wang.prl.110.093901.2013} have been explored, to name a few interesting effects.

In this paper, we investigate the influence of frequency detunings and the power of coupling fields on
the topology of the photonic band structure of honeycomb lattices that are induced in atomic vapors through the three-beam interference method.
We would like to point out that the investigation {carried out} in this paper can be extended to explore the generation of topological insulators in atomic vapor ensembles \cite{zhang.lpr.9.1.2015} and also applied to on-chip crystals,
such as praseodymium doped yttrium orthosilicate (Pr$^{3+}$:Y$_2$SiO$_5$) crystal \cite{zhang.jpcc.118.14521.2014, lan.lpl.12.015404.2015, li.rsc.2015},
which exhibits similar properties to those of atomic vapors.

The organization of the paper is as follows.
In Sec. \ref{model}, we introduce the theoretical model.
In Sec. \ref{results}, we present the results, which include the changes in
the refractive index (RI) of the material in Subsec. \ref{first} and in the corresponding PBG structures in Subsec. \ref{pbg}.
Section \ref{discussion} explores an explanation of the results and demonstrates how frequency detunings affect the topology of PBG structures.
In Sec. \ref{conclude}, we conclude the paper.

\section{The model}
\label{model}

\begin{figure}[htbp]
\centering
  \includegraphics[width=0.5\columnwidth]{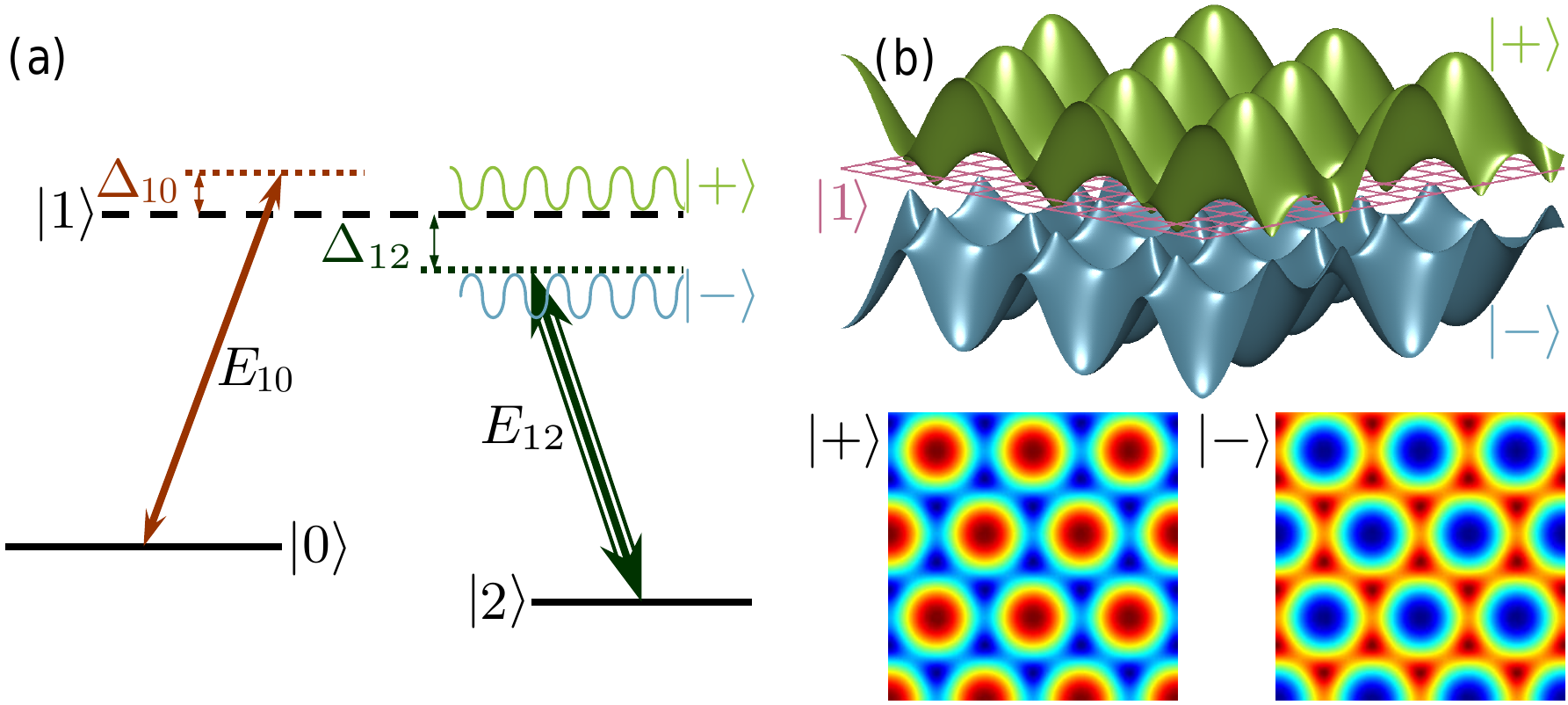}
  \caption{(Color online) (a) $\Lambda$-type energy system. (b) Energy level splitting due to the three-beam interference pattern with $\Delta_{12}>0$.
  The bottom two panels are the top views of the split sublevels $|+\rangle $ and $|-\rangle $, respectively.}
  \label{fig1}
\end{figure}

We start from a $\Lambda$-type electromagnetically induced transparency (EIT) system \cite{philips.prl.86.783.2001, artoni.prl.96.073905.2006,hang.prl.110.083604.2013,wu.light.2.e54.2013,horesley.prl.110.223602.2013}, as depicted in Fig. \ref{fig1}(a).
In the figure, the probe field $E_{10}$  connects the transition  $|0\rangle \to |1\rangle $,
and the coupling field $E_{12}$ connects the transition $|1\rangle \to |2\rangle $.
We assume that there are three broad coupling fields at the same frequency.
The three coupling fields are launched into the medium in parallel and propagate paraxially along the same direction $z$,
with the angle $2\pi/3$ between any two of them.
Therefore, they form the two-dimensional hexagonal interference pattern in the transverse $(x,y)$ plane.
Such an optically-induced hexagonal lattice can be written in the form:
\begin{equation}\label{eq1}
G = G_{12}[\exp (ik_{12}{\bf b}_1 \cdot {\bf r}) + \exp (ik_{12}{\bf b}_2 \cdot {\bf r}) + \exp (ik_{12}{\bf b}_3 \cdot {\bf r})]
\end{equation}
where ${{\bf b}_1 = ( { - {1}/{2},\, -{{\sqrt 3 }}/{2}} ) }$, ${{\bf b}_2 = ( { -{1}/{2},\,\sqrt{3}/{2}} )}$, and ${{\bf b}_3 = (1,\,0) }$.
The wave number $k_{12} $ corresponds to the coupling fields, and
$G_{12} $ represents the Rabi frequency of the coupling fields,
defined as $G_{12}=\mu_{12} E_{12}/\hbar$, with $\mu_{12}$ being the electric dipole moment.
Thus, one obtains
\[
|G|^2=|G_{12}|^2 \left[ 4\cos \left( \frac{3}{2}k_{12} x \right) \cos \bigg( \frac{\sqrt{3}}{2} k_{12} y \bigg) +  2\cos \left( \sqrt{3} k_{12} y \right) + 3 \right].
\]
Since level $|1\rangle $ is dressed by the coupling fields,
it will split into two sublevels $|+\rangle $ and $|-\rangle $
as shown by the two curves in Fig. \ref{fig1}(a), with the eigenfrequencies
\[
\ell = -\frac{1}{2}\Delta_{12} \pm \sqrt{\frac{1}{4}\Delta_{12}^2+|G|^2},
\]
according to the eigenvalue relation $\hat{H} |\pm\rangle = \hbar \ell |\pm\rangle$, where the Hamiltonian \cite{yan.pra.64.013412.2001,sun.pra.70.053820.2004,wang.pla.328.437.2004,niu.pra.84.033853.2011} is
\[
\hat{H}=-\hbar
\begin{bmatrix}
  0 & G \\
  G^\ast & \Delta_{12} \\
\end{bmatrix}.
\]
Since the interference pattern of the dressing fields is hexagonal,
the sublevels $|+\rangle $ and $|-\rangle $ split from $|1\rangle $ by such a dressing field will be periodic, as shown in Fig. \ref{fig1}(b) and in the inset panels,
in which the grid represents the original level $|1\rangle $.
One can see that the sublevels $|+\rangle $ and $|-\rangle $ exhibit hexagonal and honeycomb profiles, respectively.

Due to the periodicity of the dressing field, the susceptibility will also be periodic.
In general, the susceptibility in an atomic vapor can be written as
\begin{align}\label{eq2}
\chi_p (x,\,y)= \frac{iN\mu _{10}^2}{\hbar \epsilon_0} \frac{1}{d_{10} + |G|^2/d_{20} },
\end{align}
with $N$ being the atomic density, $\mu_{10} $ the electric dipole moment, $\epsilon_0$ the permittivity in vacuum,
and $d_{10}={\Gamma _{10}} + i{\Delta _{10}} $ and $d_{20}=\Gamma _{20} + i({\Delta _{10}} - {\Delta _{12}})$ the complex decay rates.
Here, $\Gamma_{ij} $ are the decay rates between $|i\rangle $ and $|j\rangle $ states, and
$\Delta_{10}=\Omega_{10}-\omega_{10} $ and  $\Delta_{12}=\Omega_{12}-\omega_{12} $ are the detunings.
They are determined by the transition frequencies $\Omega_{ij} $ between $|i\rangle $ and $|j\rangle $,
and by the frequencies $\omega_{10} $ and $\omega_{12} $ of the probe and the coupling fields.
The derivation of the Eq. (\ref{eq2}) can be found in the Appendix.
We note that higher-order susceptibilities should be considered when the intensity of $E_{12}$ is high enough \cite{michinel.prl.96.023903.2006,zhang.ol.37.4507.2012,zhang.pra.88.013847.2013,paredes.prl.112.173901.2014}.
However, in this paper we will stay at relatively low intensities and therefore, only consider the first-order susceptibility $\chi_p^{(1)}$.
We note that the absorption of the system, which is indicated in Eq. (\ref{eq2}) with the condition ${\Delta _{10}} - {\Delta _{12}}=0 $,
is greatly reduced. This condition corresponds to the EIT, an interesting phenomenon by which the vapor may become transparent over a narrow range \cite{harris.pt.50.36.1997,fleischhauer.rmp.77.633.2005}.

The propagation of the probe beam in the medium, in the paraxial approximation, is described by
the wave equation for the envelope of the optical electric field, which is equivalent to
the following Schr\"odinger equation \cite{rechtsman.nature.496.196.2013, plotnik.nm.13.57.2014, song.nc.6.6272.2015},
\begin{align}\label{eq4}
i\frac{{\partial \psi(x,\,y,\,z) }}{{\partial z}} + \frac{1}{{2{k_0}}}{\nabla ^2} \psi(x,\,y,\,z)
+ V_0 \frac{{{k_0}\Delta n(x,y)}}{{{n_0}}}\psi(x,\,y,\,z) = 0 ,
\end{align}
where $\psi(x,\,y,\,z) $ is the field envelope of the probe beam, $\nabla^2 = \partial^2/\partial x^2 +  \partial^2/\partial y^2 $
is the transverse Laplacian, and
$-\Delta n(x,\,y,\,z) $ in Eq. (\ref{eq4}) is the ``effective potential'' coming from the change in the refractive index,
induced by the coupling fields according to Eqs. (\ref{eq1}) and (\ref{eq2}).
Furthermore, $V_0$ is the potential depth and $k_0=2\pi n_0/\lambda $ is the wavenumber of the probe, with $\lambda$ being the wavelength.
In atomic vapors, the ambient refractive index is $n_0=\sqrt{1+{\rm{Re}}\{\chi_p(x,\,y,\,G=0)\}}$.
We should note that the minimum of the ``potential'', which is the maximum of the refractive index change,
corresponds to the lattice site.

\section{Results}
\label{results}

\subsection{Refractive index change}
\label{first}

The ``effective potential'' in Eq. (\ref{eq4}) can be written as
\begin{align}\label{deltan}
\Delta n (x,\,y) & = \sqrt {1 + {\rm Re} \{\chi_p(x,\,y)\}} - n_0 \nonumber \\
 & \approx \delta n_0 + \delta n_1 \left[ 2\cos \left(\frac{3}{2}k_{12} x \right) \cos \bigg( \frac{\sqrt{3}}{2} k_{12} y \bigg) +  \cos \left( \sqrt{3} k_{12} y \right) \right],
\end{align}
where $\delta n_0 \approx \sqrt{{\rm Re} \{1+\eta\} } - n_0$ is the spatially uniform refractive index (RI) change and
$\delta n_1 \approx -{\rm Re} \{\eta \xi\} / (2\delta n_0) $ is the coefficient of spatially-varying terms of the modulated RI. Here
$\eta=iN \mu_{10}^2 d_{20} /[\hbar \epsilon_0 (d_{10} d_{20} + 3|G_{12}|^2)] $ and $\xi \approx 2|G_{12}|^2 /(d_{10}d_{20}  + 3|G_{12}|^2) $.

\begin{figure*}[htbp]
\centering
  \includegraphics[width=\textwidth]{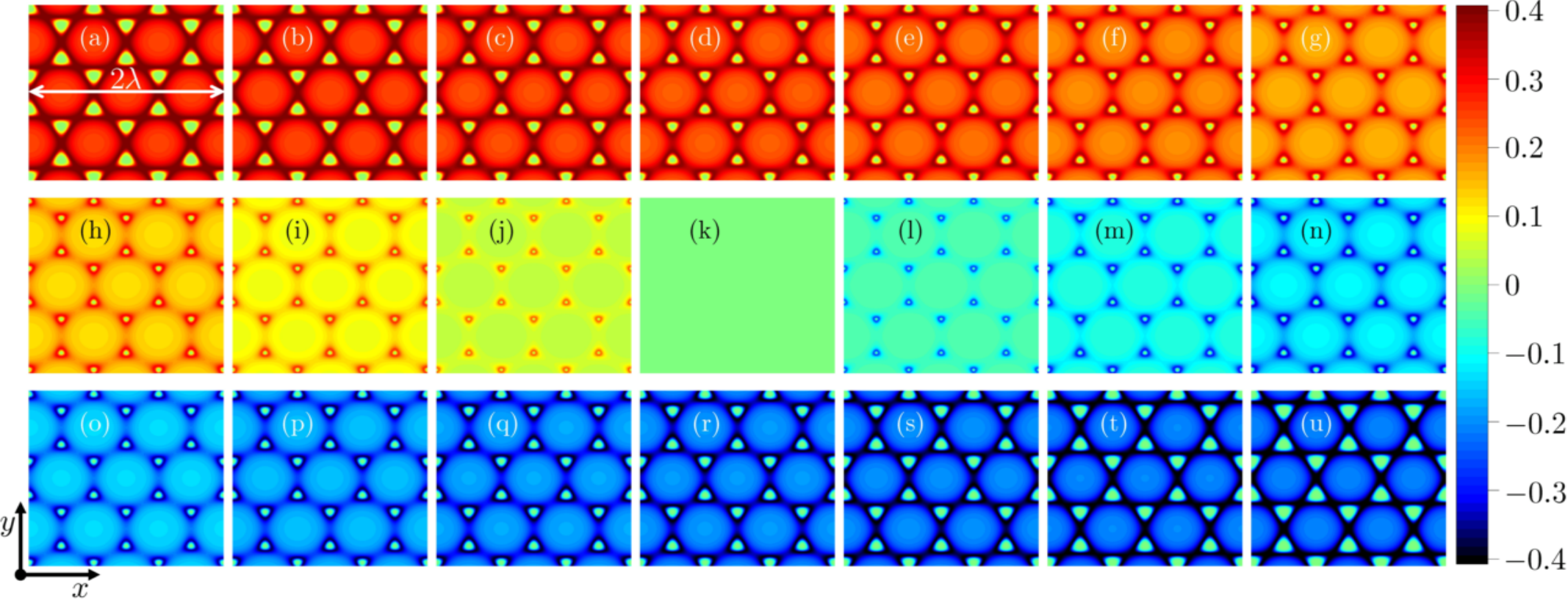}
  \caption{(Color online) Refractive index modulation $\Delta n(x,\,y) $ with $\Delta_{10}=-\Delta_{12}=-10 $ MHz (a), $-9$ MHz (b), $-8$ MHz (c), $-7$ MHz (d), $-6$ MHz (e),
  $-5$ MHz (f), $-4$ MHz (g), $-3$ MHz (h), $-2$ MHz (i), $-1$ MHz (j), 0 (k), 1 MHz (l), 2 MHz (m)
  3 MHz (n), 4 MHz (o), 5 MHz (p), 6 MHz (q), 7 MHz (r), 8 MHz (s), 9 MHz (t), and 10 MHz (u).}
  \label{fig2}
\end{figure*}

Figures \ref{fig2}(a)-\ref{fig2}(u) exhibit the RI changes for different detunings, as displayed in the caption.
In Figs. \ref{fig2}(a)-\ref{fig2}(j), with the frequency detuning $\Delta_{10}<0$, the induced RI change at the honeycomb lattice sites is the smallest,
but in the regions immediately around the sites it is the biggest, therefore the induced RI change exhibits a {\color{red}honeycomb-like} pattern.
When $\Delta_{10}>0$, as shown in Figs. \ref{fig2}(l)-\ref{fig2}(u), the RI change at the lattice sites is the biggest and
in the regions around the sites it is the smallest, which also exhibits a honeycomb pattern.
If $\Delta_{10}=\Delta_{12}=0 $, the linear susceptibility in Eqs. (\ref{eq1}) and (\ref{eq2}) is imaginary
so that RI is always a constant in the transverse plane and the honeycomb lattice disappears, according to Eq. (\ref{deltan}), which is shown in Fig. \ref{fig2}(k).

\subsection{Topology of the PBG structure}
\label{pbg}

\begin{figure*}[htbp]
\centering
  \includegraphics[width=\textwidth]{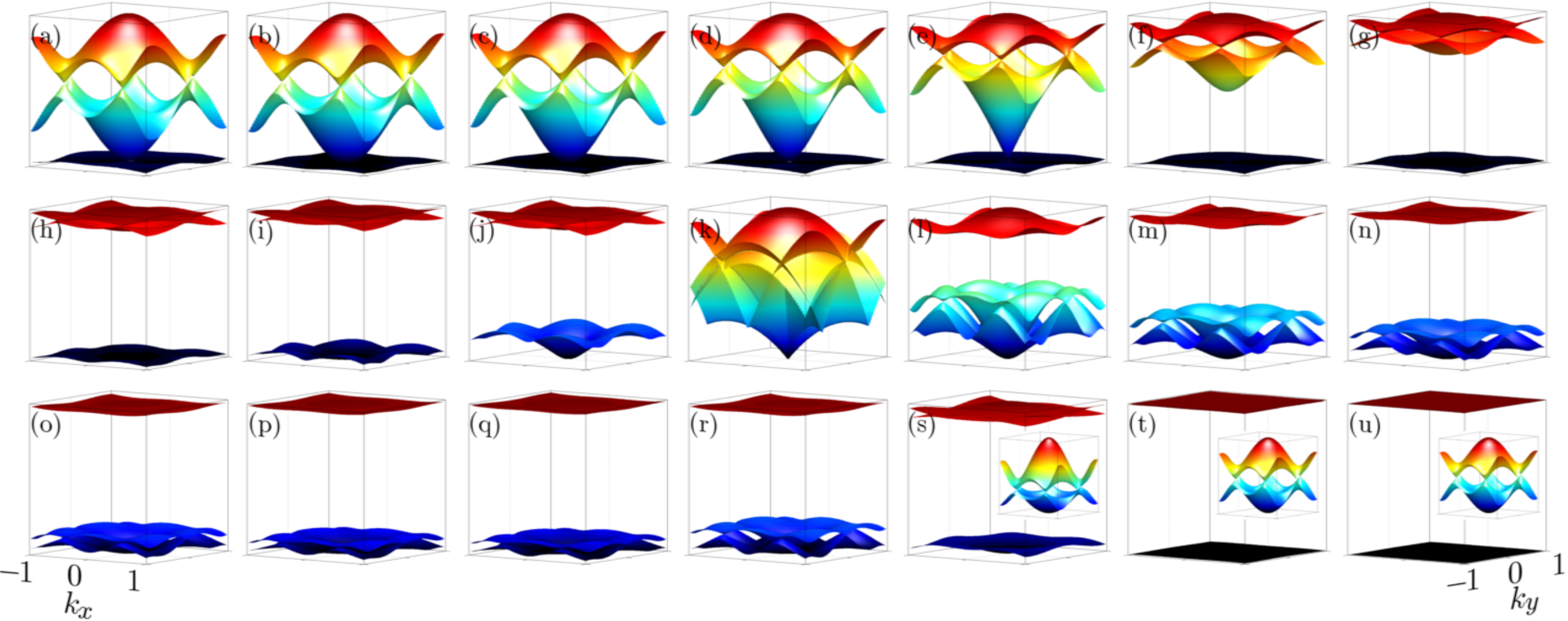}
  \caption{(Color online) Photonic band structures corresponding to Fig. \ref{fig2}.
  Panel (k) presents a zoomed-in band structure of the flat RI from Fig. \ref{fig2}(k).
  Insets in (s)-(u) depict similar zoomed-in structures.}
  \label{fig3}
\end{figure*}

Corresponding to the RI changes in Fig. \ref{fig2}, the PBG structure in the first Brillouin zone is shown in Fig. \ref{fig3},
based on the plane wave expansion method.
In each panel of Fig. \ref{fig3}, we display the lowest three bands.
In Figs. \ref{fig3}(a)-\ref{fig3}(j), one can see that the PBG structure contains 6 Dirac cones (between the upper two bands) at the vertices of the first Brillouin zone,
and the dispersion relation is linear around the Dirac points.
Corresponding to Fig. \ref{fig2}(k), the medium is homogeneous and the PBG structure is shown in Fig. \ref{fig3}(k),
in which the edges of the upper two bands merge with each other and Dirac cones disappear in the PBG.
However, the PBG structures shown in Figs. \ref{fig3}(l)-\ref{fig3}(r), which correspond to Figs. \ref{fig2}(l)-\ref{fig2}(r),
are quite different from those in Figs. \ref{fig3}(a)-\ref{fig3}(j).
In Figs. \ref{fig3}(l)-\ref{fig3}(r), a big band gap opens between the upper two bands, so that the Dirac cones disappear.
When the frequency detuning increases further, as shown in Figs. \ref{fig3}(s)-\ref{fig3}(u),
the upper two bands become almost degenerate and flat, and there is a big band gap between the upper two bands and the bottom band.
When the upper two bands in Figs. \ref{fig3}(s)-\ref{fig3}(u) are zoomed in, as shown in the insets,
one can find that there are still 6 Dirac points at the vertices of the first Brillouin zone.
One has to take into account that the difference in height of the upper two bands is too small to recognize the cones,
in comparison with the big band gap.
Therefore, even if the edges of the upper two bands seemingly merge with each other,
there are still Dirac cones visible \cite{ablowitz.jam.72.240.2012} at the 6 corners of the first Brillouin zone.
In other words, even though $\Delta_{10}>0$ in Figs. \ref{fig3}(l)-\ref{fig3}(u),
the PBG topology in Figs. \ref{fig3}(s)-\ref{fig3}(u) is different from those in Figs. \ref{fig3}(l)-\ref{fig3}(r),
and is similar to some extent to the cases in Figs. \ref{fig3}(f)-\ref{fig3}(j), where $\Delta_{10}<0$.

\begin{figure*}[htbp]
\centering
  \includegraphics[width=\textwidth]{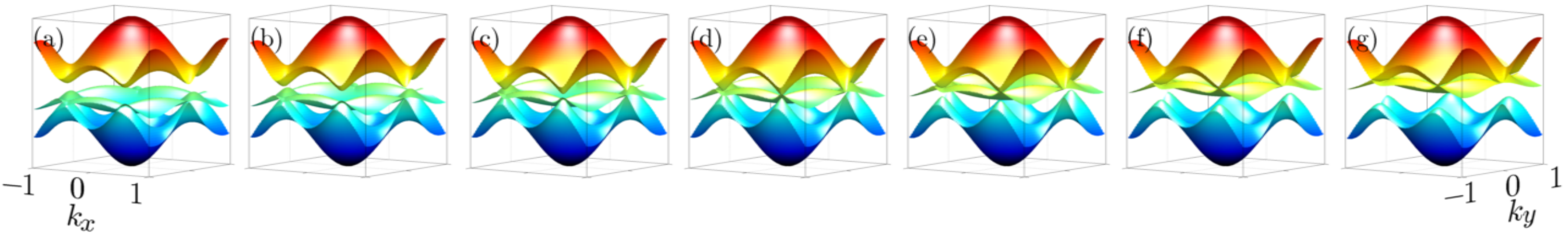}
  \caption{(Color online) Photonic band structures corresponding to the cases with $\Delta_{10}=-\Delta_{12}=7.63 $ MHz (a), $7.64$ MHz (b), $7.65$ MHz (c), $7.66$ MHz (d), $7.67$ MHz (e), $7.68$ MHz (f), and $7.69$ MHz (g), respectively.}
  \label{fig4}
\end{figure*}

Figure \ref{fig3} clearly exhibits the influence of frequency detunings on the topology of the PBG structure,
especially in Figs. \ref{fig3}(r) and \ref{fig3}(s).
In order to perceive evolution of the topology of the PBG structure, we also display the cases in-between Figs. \ref{fig3}(r) and \ref{fig3}(s), as shown in Fig. \ref{fig4}.
We would like to note that even though the transition between Figs. \ref{fig3}(j)-\ref{fig3}(l) is quite abrupt,
the change is continuous, so it does not provide additional information. Hence, we do not show the intermediate stages.
At the 6 corners of the first Brillouin zone, the gaps between the upper two bands decrease and then merge together,
while the opposite is  observed for the bottom two bands.
One finds that such a ``transition'' happens at a certain detuning -- the one shown in Fig. \ref{fig4}(d).

\section{Discussion}
\label{discussion}

What is the physical reason behind such a PBG modulation? In this section, we answer this question.
Since the PBG structure reflects the property of the changing RI, it is natural to go back to the RI change and find the explanation.

In order to see the trend in the RI change with the frequency detuning more clearly, we display the RI change in the $x=0$ plane versus $\Delta_{10}$ in Fig. \ref{fig5}(a),
while in Figs. \ref{fig5}(b) and \ref{fig5}(c), we show the RI change in the $x=0$ plane corresponding to Figs. \ref{fig2}(f) and \ref{fig2}(p).
As stated in the previous literature \cite{rechtsman.nature.496.196.2013, plotnik.nm.13.57.2014, song.nc.6.6272.2015}, a honeycomb lattice demands the RI at the lattice sites to be bigger than the ambient RI.
However, a more complex situation is encountered around honeycomb sites in an atomic vapor. Consequently, the cases with $\Delta_{10}<0$ and $\Delta_{12}>0$ have to be discussed separately. This is done in the following two subsections.

\subsection{The case $\Delta_{10}<0$}

As mentioned, the RI at the sites is then the smallest.
However, the RI immediately around the sites is the biggest, and it forms a honeycomb lattice.
In other words, the honeycomb lattice is composed of small triangular structures -- one can observe such lattices in Figs. \ref{fig2}(a)-\ref{fig2}(j).
For this case, the honeycomb lattice sites should properly refer to these triangular structures.
In Figs. \ref{fig5}(a) and \ref{fig5}(b), one can see the distance between the two nearest neighbor (NN) sites (two peaks in Fig. \ref{fig5}(b)),
and this distance cannot block the hopping between the two sites;
therefore one can observe the Dirac cones in the PBG structure.

With the increasing frequency detuning, the distance increases and broadens, which makes the hopping more difficult;
as a result, the topology of the PBG structure is affected -- the heights of the upper two bands shrink, even though there are still Dirac points.

\subsection{The case  $\Delta_{10}>0$}

\begin{figure}[htbp]
\centering
  \includegraphics[width=0.5\columnwidth]{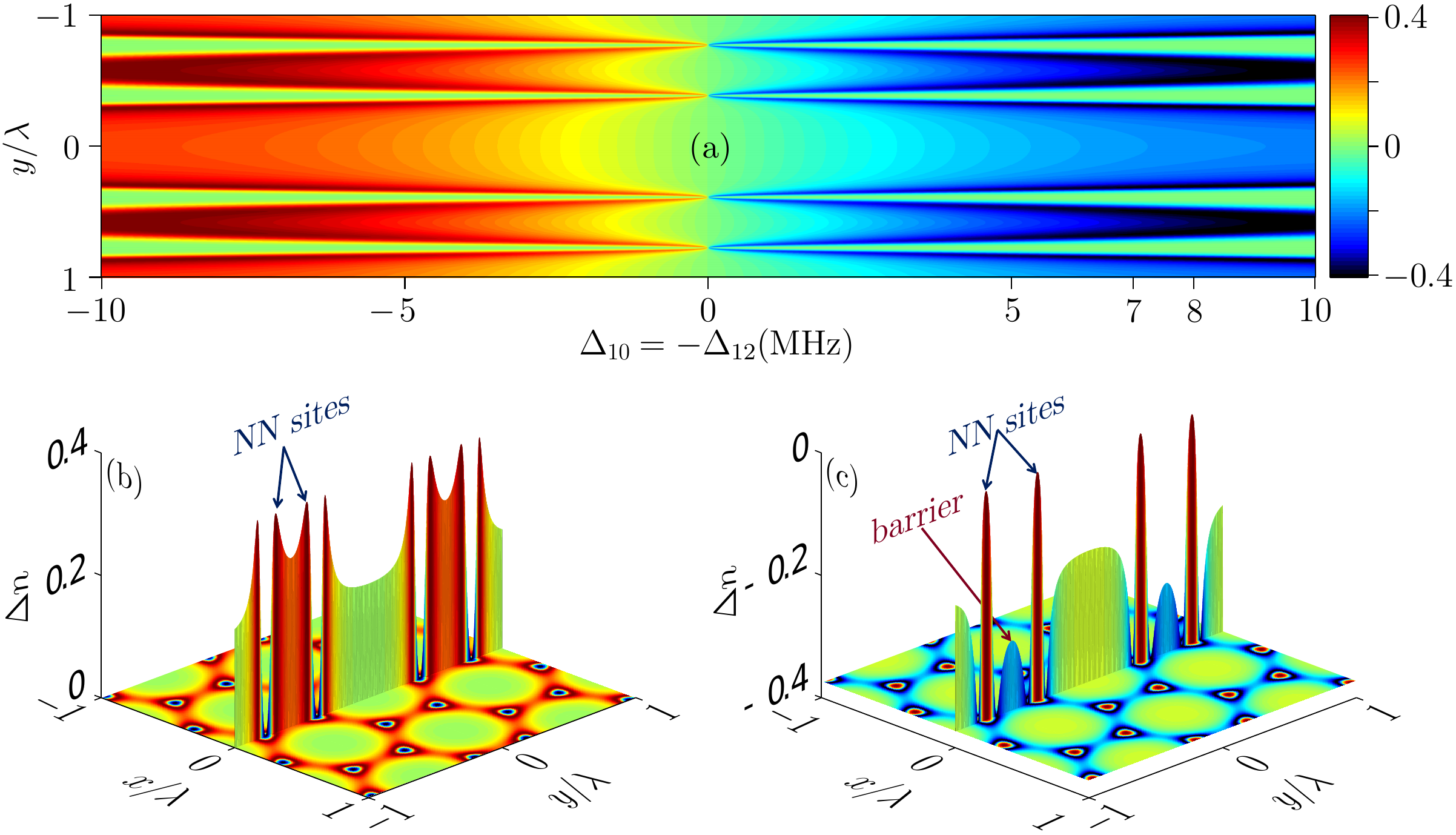}
  \caption{(Color online) (a) RI change along the cross section $x=0$ versus different frequency detunings.
  (b) RI change corresponding to Fig. \ref{fig2}(f), as a color plot in the $xy$ plane, with $\Delta n$ along the $x = 0$ line plotted explicitly on the vertical axis.
  (c) Same as (b), but corresponding to Fig. \ref{fig2}(p).}
  \label{fig5}
\end{figure}

When $\Delta_{10}>0$, the RI change at the sites is the biggest; therefore, the sites directly form a honeycomb lattice,
as shown in Figs. \ref{fig2}(l)-\ref{fig2}(u), which is substantially different from the $\Delta_{10}<0$ case.
From Figs. \ref{fig5}(a) and \ref{fig5}(c), one can observe the appearance of a ``potential barrier'' between the two NN sites.
Such a barrier does not exist in the case $\Delta_{10}<0$, because the RI change is always the highest around the sites,
and the RI change between the two peaks, as shown in Fig. \ref{fig5}(b), monotonously decreases and then monotonously increases along the positive $y$ direction (that is, there is no potential barrier).

The existence of the barrier blocks the hopping effectively,
so that no Dirac cone forms in the PBG structure.
We call this phenomenon in the PBG topology the ``\textit{blockage effect}''.
With an increasing frequency detuning the barrier decreases, and the blockage effect also weakens.
Numerical simulation demonstrates that the blockage effect disappears at about $\Delta_{10}=-\Delta_{12} \approx 7.66$ MHz, as displayed in Fig. \ref{fig4}(d).
Without the blockage effect, the hopping between NN sites happens, and correspondingly there appear Dirac cones between the upper two bands in the PBG structure,
as shown in the insets of Figs. \ref{fig3}(s)-\ref{fig3}(u).
Since the RI change at the sites is much bigger than the ambient RI change, the hopping strength is relatively weak.
This is the reason why the upper two bands in Figs. \ref{fig3}(s)-\ref{fig3}(u) are almost degenerate.

\section{Conclusion}
\label{conclude}

In summary, we have investigated the influence of changing frequency detunings on the topology of the PBG structure of the honeycomb  lattice induced in an atomic vapor.
We have illuminated how the appearance and disappearance of Dirac points and cones is associated with the change in the sign of the frequency detuning.
This research can be utilized in the construction
of topological insulators, but also can be applied to the Pr$^{3+}$:Y$_2$SiO$_5$
crystal, even though it is developed for gaseous systems.
Our investigation broadens the understanding of PBG structures of the honeycomb  lattice,
and may help in fabricating photonic devices that can be used in all-optical signal processing and computing.

\section*{Acknowledgments}
This work was supported by the 973 Program (2012CB921804),
NSFC (61308015, 11474228), KSTIT of Shaanxi Province (2014KCT-10),
and NPRP 6-021-1-005 project from the Qatar National Research Fund (a member of the Qatar Foundation).

\section*{Appendix: Derivation of susceptibilities in Eq. (\ref{eq2})}

The master equations are:
\begin{subequations}
\begin{equation}\label{eqs1a}
\frac{\partial }{{\partial t}}\rho _{00}^{(r)}  =  - \Gamma _{00} \rho _{00}^{(r)}  + i[G_{10}^* \rho _{10}^{(r - 1)}  - G_{10} \rho _{01}^{(r - 1)}],\tag{A.1a}
\end{equation}
\begin{equation}\label{eqs1b}
\frac{\partial }{{\partial t}}\rho _{11}^{(r)}  =  - \Gamma _{11} \rho _{11}^{(r)}  + i[(G_{10} \rho _{01}^{(r - 1)}  - G_{10}^* \rho _{10}^{(r - 1)} ) + G_{12}^* (\rho _{21}^{(r - 1)}  - \rho _{12}^{(r - 1)} )],\tag{A.1b}
\end{equation}
\begin{equation}\label{eqs1c}
\frac{\partial }{{\partial t}}\rho _{10}^{(r)}  =  - (i\Delta _{10}  + \Gamma _{10} )\rho _{10}^{(r)}  + i[G_{10} (\rho _{00}^{(r - 1)}  - \rho _{11}^{(r - 1)} ) + G_{12}^* \rho _{20}^{(r - 1)} ],\tag{A.1c}
\end{equation}
\begin{equation}\label{eqs1d}
\frac{\partial }{{\partial t}}\rho _{20}^{(r)}  =  - [i(\Delta _{10}  - \Delta _{12} ) + \Gamma _{20} ]\rho _{20}^{(r)}  + i[G_{12}^* \rho _{10}^{(r - 1)}  - G_{10} \rho _{21}^{(r - 1)} ],\tag{A.1d}
\end{equation}
\begin{equation}\label{eqs1e}
\frac{\partial }{{\partial t}}\rho _{21}^{(r)}  =  - (\Gamma _{21}  - i\Delta _{12} )\rho _{21}^{(r)}  + i[G_{12} (\rho _{11}^{(r - 1)}  - \rho _{22}^{(r - 1)} ) - G_{10}^* \rho _{20}^{(r - 1)} ].\tag{A.1e}
\end{equation}
\end{subequations}

A photon is driven from $|0\rangle $ to $|1\rangle $ by $E_{10}$,
and the corresponding perturbation chain is $\rho _{00}^{(0)} \xrightarrow{G_{10}} \rho _{G_{12} \pm 0}^{(1)}$ \cite{zhang.book.2009}.
We assume that $\rho _{00}^{(0)}  \approx 1$, $\rho _{11}^{(0)}  \approx 0 $, and $\rho _{22}^{(0)}  \approx 0 $.
Therefore, Eqs. (\ref{eqs1c}) and (\ref{eqs1d}) can be rewritten as
\begin{subequations}
\begin{equation}\label{eqs2a}
\frac{\partial }{{\partial t}}\rho _{G_{12} \pm 0}  =  - (i\Delta _{10}  + \Gamma _{10} )\rho _{G_{12} \pm 0}  + i(G_{10} \rho _{00}  + G_{12}^* \rho _{20} ),\tag{A.2a}
\end{equation}
\begin{equation}\label{eqs2b}
\frac{\partial }{{\partial t}}\rho _{20}  =  - [i(\Delta _{10}  - \Delta _{12} ) + \Gamma _{20} ]\rho _{20}  + iG_{12}^* \rho _{G_{12} \pm 0} .\tag{A.2b}
\end{equation}
\end{subequations}
Based on Eqs. (\ref{eqs2a}) and (\ref{eqs2b}), one obtains
\begin{equation}\label{eqs3}
\rho _{G_{12} \pm 0}^{(1)}  = \frac{{iG_{10} }}{{{\Gamma _{10}  + i\Delta _{10}  + \frac{{\left| {G_{12} } \right|^2 }}{{\Gamma _{20}  + i\Delta _{10}  - i\Delta _{12} }}} }}\rho _{00}^{(0)}.\tag{A.3}
\end{equation}
According to the relation  $\epsilon _0 \chi ^{(1)} E_{10}  = N\mu_{10} \rho _{10}^{(1)}  $ \cite{wu.josab.25.1840.2008}, one finds
\begin{equation}\label{eqs4}
\chi ^{(1)}  = \frac{iN\mu _{10}^2}{\hbar \epsilon _0} \frac{1}{\Gamma _{10}  + i\Delta _{10}  + \frac{|G_{12}|^2}{\Gamma _{20}  + i\Delta _{10}  - i\Delta _{12}}}.\tag{A.4}
\end{equation}
Equation (\ref{eqs4}) is just the susceptibility used in Eq. (\ref{eq2}).


\begin{thebibliography}{46}%
\makeatletter
\providecommand \@ifxundefined [1]{%
 \@ifx{#1\undefined}
}%
\providecommand \@ifnum [1]{%
 \ifnum #1\expandafter \@firstoftwo
 \else \expandafter \@secondoftwo
 \fi
}%
\providecommand \@ifx [1]{%
 \ifx #1\expandafter \@firstoftwo
 \else \expandafter \@secondoftwo
 \fi
}%
\providecommand \natexlab [1]{#1}%
\providecommand \enquote  [1]{``#1''}%
\providecommand \bibnamefont  [1]{#1}%
\providecommand \bibfnamefont [1]{#1}%
\providecommand \citenamefont [1]{#1}%
\providecommand \href@noop [0]{\@secondoftwo}%
\providecommand \href [0]{\begingroup \@sanitize@url \@href}%
\providecommand \@href[1]{\@@startlink{#1}\@@href}%
\providecommand \@@href[1]{\endgroup#1\@@endlink}%
\providecommand \@sanitize@url [0]{\catcode `\\12\catcode `\$12\catcode
  `\&12\catcode `\#12\catcode `\^12\catcode `\_12\catcode `\%12\relax}%
\providecommand \@@startlink[1]{}%
\providecommand \@@endlink[0]{}%
\providecommand \url  [0]{\begingroup\@sanitize@url \@url }%
\providecommand \@url [1]{\endgroup\@href {#1}{\urlprefix }}%
\providecommand \urlprefix  [0]{URL }%
\providecommand \Eprint [0]{\href }%
\providecommand \doibase [0]{http://dx.doi.org/}%
\providecommand \selectlanguage [0]{\@gobble}%
\providecommand \bibinfo  [0]{\@secondoftwo}%
\providecommand \bibfield  [0]{\@secondoftwo}%
\providecommand \translation [1]{[#1]}%
\providecommand \BibitemOpen [0]{}%
\providecommand \bibitemStop [0]{}%
\providecommand \bibitemNoStop [0]{.\EOS\space}%
\providecommand \EOS [0]{\spacefactor3000\relax}%
\providecommand \BibitemShut  [1]{\csname bibitem#1\endcsname}%
\let\auto@bib@innerbib\@empty
\bibitem [{\citenamefont {Peleg}\ \emph {et~al.}(2007)\citenamefont {Peleg},
  \citenamefont {Bartal}, \citenamefont {Freedman}, \citenamefont {Manela},
  \citenamefont {Segev},\ and\ \citenamefont
  {Christodoulides}}]{peleg.prl.98.103901.2007}%
  \BibitemOpen
  \bibfield  {author} {\bibinfo {author} {\bibfnamefont {O.}~\bibnamefont
  {Peleg}}, \bibinfo {author} {\bibfnamefont {G.}~\bibnamefont {Bartal}},
  \bibinfo {author} {\bibfnamefont {B.}~\bibnamefont {Freedman}}, \bibinfo
  {author} {\bibfnamefont {O.}~\bibnamefont {Manela}}, \bibinfo {author}
  {\bibfnamefont {M.}~\bibnamefont {Segev}}, \ and\ \bibinfo {author}
  {\bibfnamefont {D.~N.}\ \bibnamefont {Christodoulides}},\ }\href {\doibase
  10.1103/PhysRevLett.98.103901} {\bibfield  {journal} {\bibinfo  {journal}
  {Phys. Rev. Lett.}\ }\textbf {\bibinfo {volume} {98}},\ \bibinfo {pages}
  {103901} (\bibinfo {year} {2007})}\BibitemShut {NoStop}%
\bibitem [{\citenamefont {Ablowitz}\ \emph {et~al.}(2009)\citenamefont
  {Ablowitz}, \citenamefont {Nixon},\ and\ \citenamefont
  {Zhu}}]{ablowitz.pra.79.053830.2009}%
  \BibitemOpen
  \bibfield  {author} {\bibinfo {author} {\bibfnamefont {M.~J.}\ \bibnamefont
  {Ablowitz}}, \bibinfo {author} {\bibfnamefont {S.~D.}\ \bibnamefont {Nixon}},
  \ and\ \bibinfo {author} {\bibfnamefont {Y.}~\bibnamefont {Zhu}},\ }\href
  {\doibase 10.1103/PhysRevA.79.053830} {\bibfield  {journal} {\bibinfo
  {journal} {Phys. Rev. A}\ }\textbf {\bibinfo {volume} {79}},\ \bibinfo
  {pages} {053830} (\bibinfo {year} {2009})}\BibitemShut {NoStop}%
\bibitem [{\citenamefont {Bahat-Treidel}\ \emph {et~al.}(2010)\citenamefont
  {Bahat-Treidel}, \citenamefont {Peleg}, \citenamefont {Grobman},
  \citenamefont {Shapira}, \citenamefont {Segev},\ and\ \citenamefont
  {Pereg-Barnea}}]{bahat-treidel.prl.104.063901.2010}%
  \BibitemOpen
  \bibfield  {author} {\bibinfo {author} {\bibfnamefont {O.}~\bibnamefont
  {Bahat-Treidel}}, \bibinfo {author} {\bibfnamefont {O.}~\bibnamefont
  {Peleg}}, \bibinfo {author} {\bibfnamefont {M.}~\bibnamefont {Grobman}},
  \bibinfo {author} {\bibfnamefont {N.}~\bibnamefont {Shapira}}, \bibinfo
  {author} {\bibfnamefont {M.}~\bibnamefont {Segev}}, \ and\ \bibinfo {author}
  {\bibfnamefont {T.}~\bibnamefont {Pereg-Barnea}},\ }\href {\doibase
  10.1103/PhysRevLett.104.063901} {\bibfield  {journal} {\bibinfo  {journal}
  {Phys. Rev. Lett.}\ }\textbf {\bibinfo {volume} {104}},\ \bibinfo {pages}
  {063901} (\bibinfo {year} {2010})}\BibitemShut {NoStop}%
\bibitem [{\citenamefont {Terhalle}\ \emph {et~al.}(2008)\citenamefont
  {Terhalle}, \citenamefont {Richter}, \citenamefont {Desyatnikov},
  \citenamefont {Neshev}, \citenamefont {Krolikowski}, \citenamefont {Kaiser},
  \citenamefont {Denz},\ and\ \citenamefont
  {Kivshar}}]{terhalle.prl.101.013903.2008}%
  \BibitemOpen
  \bibfield  {author} {\bibinfo {author} {\bibfnamefont {B.}~\bibnamefont
  {Terhalle}}, \bibinfo {author} {\bibfnamefont {T.}~\bibnamefont {Richter}},
  \bibinfo {author} {\bibfnamefont {A.~S.}\ \bibnamefont {Desyatnikov}},
  \bibinfo {author} {\bibfnamefont {D.~N.}\ \bibnamefont {Neshev}}, \bibinfo
  {author} {\bibfnamefont {W.}~\bibnamefont {Krolikowski}}, \bibinfo {author}
  {\bibfnamefont {F.}~\bibnamefont {Kaiser}}, \bibinfo {author} {\bibfnamefont
  {C.}~\bibnamefont {Denz}}, \ and\ \bibinfo {author} {\bibfnamefont {Y.~S.}\
  \bibnamefont {Kivshar}},\ }\href {\doibase 10.1103/PhysRevLett.101.013903}
  {\bibfield  {journal} {\bibinfo  {journal} {Phys. Rev. Lett.}\ }\textbf
  {\bibinfo {volume} {101}},\ \bibinfo {pages} {013903} (\bibinfo {year}
  {2008})}\BibitemShut {NoStop}%
\bibitem [{\citenamefont {Terhalle}\ \emph {et~al.}(2011)\citenamefont
  {Terhalle}, \citenamefont {Desyatnikov}, \citenamefont {Neshev},
  \citenamefont {Krolikowski}, \citenamefont {Denz},\ and\ \citenamefont
  {Kivshar}}]{terhalle.prl.106.083902.2011}%
  \BibitemOpen
  \bibfield  {author} {\bibinfo {author} {\bibfnamefont {B.}~\bibnamefont
  {Terhalle}}, \bibinfo {author} {\bibfnamefont {A.~S.}\ \bibnamefont
  {Desyatnikov}}, \bibinfo {author} {\bibfnamefont {D.~N.}\ \bibnamefont
  {Neshev}}, \bibinfo {author} {\bibfnamefont {W.}~\bibnamefont {Krolikowski}},
  \bibinfo {author} {\bibfnamefont {C.}~\bibnamefont {Denz}}, \ and\ \bibinfo
  {author} {\bibfnamefont {Y.~S.}\ \bibnamefont {Kivshar}},\ }\href {\doibase
  10.1103/PhysRevLett.106.083902} {\bibfield  {journal} {\bibinfo  {journal}
  {Phys. Rev. Lett.}\ }\textbf {\bibinfo {volume} {106}},\ \bibinfo {pages}
  {083902} (\bibinfo {year} {2011})}\BibitemShut {NoStop}%
\bibitem [{\citenamefont {Tarruell}\ \emph {et~al.}(2012)\citenamefont
  {Tarruell}, \citenamefont {Greif}, \citenamefont {Uehlinger}, \citenamefont
  {Jotzu},\ and\ \citenamefont {Esslinger}}]{tarruell.nature.483.302.2012}%
  \BibitemOpen
  \bibfield  {author} {\bibinfo {author} {\bibfnamefont {L.}~\bibnamefont
  {Tarruell}}, \bibinfo {author} {\bibfnamefont {D.}~\bibnamefont {Greif}},
  \bibinfo {author} {\bibfnamefont {T.}~\bibnamefont {Uehlinger}}, \bibinfo
  {author} {\bibfnamefont {G.}~\bibnamefont {Jotzu}}, \ and\ \bibinfo {author}
  {\bibfnamefont {T.}~\bibnamefont {Esslinger}},\ }\href {\doibase
  10.1038/nature10871} {\bibfield  {journal} {\bibinfo  {journal} {Nature}\
  }\textbf {\bibinfo {volume} {483}},\ \bibinfo {pages} {302} (\bibinfo {year}
  {2012})}\BibitemShut {NoStop}%
\bibitem [{\citenamefont {Gomes}\ \emph {et~al.}(2012)\citenamefont {Gomes},
  \citenamefont {Mar}, \citenamefont {Ko}, \citenamefont {Guinea},\ and\
  \citenamefont {Manoharan}}]{gomes.nature.483.306.2012}%
  \BibitemOpen
  \bibfield  {author} {\bibinfo {author} {\bibfnamefont {K.~K.}\ \bibnamefont
  {Gomes}}, \bibinfo {author} {\bibfnamefont {W.}~\bibnamefont {Mar}}, \bibinfo
  {author} {\bibfnamefont {W.}~\bibnamefont {Ko}}, \bibinfo {author}
  {\bibfnamefont {F.}~\bibnamefont {Guinea}}, \ and\ \bibinfo {author}
  {\bibfnamefont {H.~C.}\ \bibnamefont {Manoharan}},\ }\href {\doibase
  10.1038/nature10941} {\bibfield  {journal} {\bibinfo  {journal} {Nature}\
  }\textbf {\bibinfo {volume} {483}},\ \bibinfo {pages} {306} (\bibinfo {year}
  {2012})}\BibitemShut {NoStop}%
\bibitem [{\citenamefont {Rechtsman}\ \emph
  {et~al.}(2013{\natexlab{a}})\citenamefont {Rechtsman}, \citenamefont
  {Zeuner}, \citenamefont {Plotnik}, \citenamefont {Lumer}, \citenamefont
  {Podolsky}, \citenamefont {Dreisow}, \citenamefont {Nolte}, \citenamefont
  {Segev},\ and\ \citenamefont {Szameit}}]{rechtsman.nature.496.196.2013}%
  \BibitemOpen
  \bibfield  {author} {\bibinfo {author} {\bibfnamefont {M.~C.}\ \bibnamefont
  {Rechtsman}}, \bibinfo {author} {\bibfnamefont {J.~M.}\ \bibnamefont
  {Zeuner}}, \bibinfo {author} {\bibfnamefont {Y.}~\bibnamefont {Plotnik}},
  \bibinfo {author} {\bibfnamefont {Y.}~\bibnamefont {Lumer}}, \bibinfo
  {author} {\bibfnamefont {D.}~\bibnamefont {Podolsky}}, \bibinfo {author}
  {\bibfnamefont {F.}~\bibnamefont {Dreisow}}, \bibinfo {author} {\bibfnamefont
  {S.}~\bibnamefont {Nolte}}, \bibinfo {author} {\bibfnamefont
  {M.}~\bibnamefont {Segev}}, \ and\ \bibinfo {author} {\bibfnamefont
  {A.}~\bibnamefont {Szameit}},\ }\href {\doibase 10.1038/nature12066}
  {\bibfield  {journal} {\bibinfo  {journal} {Nature}\ }\textbf {\bibinfo
  {volume} {496}},\ \bibinfo {pages} {196} (\bibinfo {year}
  {2013}{\natexlab{a}})}\BibitemShut {NoStop}%
\bibitem [{\citenamefont {Rechtsman}\ \emph
  {et~al.}(2013{\natexlab{b}})\citenamefont {Rechtsman}, \citenamefont
  {Zeuner}, \citenamefont {Tunnermann}, \citenamefont {Nolte}, \citenamefont
  {Segev},\ and\ \citenamefont {Szameit}}]{rechtsman.np.7.153.2013}%
  \BibitemOpen
  \bibfield  {author} {\bibinfo {author} {\bibfnamefont {M.~C.}\ \bibnamefont
  {Rechtsman}}, \bibinfo {author} {\bibfnamefont {J.~M.}\ \bibnamefont
  {Zeuner}}, \bibinfo {author} {\bibfnamefont {A.}~\bibnamefont {Tunnermann}},
  \bibinfo {author} {\bibfnamefont {S.}~\bibnamefont {Nolte}}, \bibinfo
  {author} {\bibfnamefont {M.}~\bibnamefont {Segev}}, \ and\ \bibinfo {author}
  {\bibfnamefont {A.}~\bibnamefont {Szameit}},\ }\href {\doibase
  10.1038/nphoton.2012.302} {\bibfield  {journal} {\bibinfo  {journal} {Nat.
  Photon.}\ }\textbf {\bibinfo {volume} {7}},\ \bibinfo {pages} {153} (\bibinfo
  {year} {2013}{\natexlab{b}})}\BibitemShut {NoStop}%
\bibitem [{\citenamefont {Rechtsman}\ \emph
  {et~al.}(2013{\natexlab{c}})\citenamefont {Rechtsman}, \citenamefont
  {Plotnik}, \citenamefont {Zeuner}, \citenamefont {Song}, \citenamefont
  {Chen}, \citenamefont {Szameit},\ and\ \citenamefont
  {Segev}}]{rechtsman.prl.111.103901.2013}%
  \BibitemOpen
  \bibfield  {author} {\bibinfo {author} {\bibfnamefont {M.~C.}\ \bibnamefont
  {Rechtsman}}, \bibinfo {author} {\bibfnamefont {Y.}~\bibnamefont {Plotnik}},
  \bibinfo {author} {\bibfnamefont {J.~M.}\ \bibnamefont {Zeuner}}, \bibinfo
  {author} {\bibfnamefont {D.}~\bibnamefont {Song}}, \bibinfo {author}
  {\bibfnamefont {Z.}~\bibnamefont {Chen}}, \bibinfo {author} {\bibfnamefont
  {A.}~\bibnamefont {Szameit}}, \ and\ \bibinfo {author} {\bibfnamefont
  {M.}~\bibnamefont {Segev}},\ }\href {\doibase 10.1103/PhysRevLett.111.103901}
  {\bibfield  {journal} {\bibinfo  {journal} {Phys. Rev. Lett.}\ }\textbf
  {\bibinfo {volume} {111}},\ \bibinfo {pages} {103901} (\bibinfo {year}
  {2013}{\natexlab{c}})}\BibitemShut {NoStop}%
\bibitem [{\citenamefont {Plotnik}\ \emph {et~al.}(2014)\citenamefont
  {Plotnik}, \citenamefont {Rechtsman}, \citenamefont {Song}, \citenamefont
  {Heinrich}, \citenamefont {Zeuner}, \citenamefont {Nolte}, \citenamefont
  {Lumer}, \citenamefont {Malkova}, \citenamefont {Xu}, \citenamefont
  {Szameit}, \citenamefont {Chen},\ and\ \citenamefont
  {Segev}}]{plotnik.nm.13.57.2014}%
  \BibitemOpen
  \bibfield  {author} {\bibinfo {author} {\bibfnamefont {Y.}~\bibnamefont
  {Plotnik}}, \bibinfo {author} {\bibfnamefont {M.~C.}\ \bibnamefont
  {Rechtsman}}, \bibinfo {author} {\bibfnamefont {D.}~\bibnamefont {Song}},
  \bibinfo {author} {\bibfnamefont {M.}~\bibnamefont {Heinrich}}, \bibinfo
  {author} {\bibfnamefont {J.~M.}\ \bibnamefont {Zeuner}}, \bibinfo {author}
  {\bibfnamefont {S.}~\bibnamefont {Nolte}}, \bibinfo {author} {\bibfnamefont
  {Y.}~\bibnamefont {Lumer}}, \bibinfo {author} {\bibfnamefont
  {N.}~\bibnamefont {Malkova}}, \bibinfo {author} {\bibfnamefont
  {J.}~\bibnamefont {Xu}}, \bibinfo {author} {\bibfnamefont {A.}~\bibnamefont
  {Szameit}}, \bibinfo {author} {\bibfnamefont {Z.~C.}\ \bibnamefont {Chen}}, \
  and\ \bibinfo {author} {\bibfnamefont {M.}~\bibnamefont {Segev}},\ }\href
  {\doibase 10.1038/nmat3783} {\bibfield  {journal} {\bibinfo  {journal} {Nat.
  Mater.}\ }\textbf {\bibinfo {volume} {13}},\ \bibinfo {pages} {57} (\bibinfo
  {year} {2014})}\BibitemShut {NoStop}%
\bibitem [{\citenamefont {Jotzu}\ \emph {et~al.}(2014)\citenamefont {Jotzu},
  \citenamefont {Messer}, \citenamefont {Desbuquois}, \citenamefont {Lebrat},
  \citenamefont {Uehlinger}, \citenamefont {Greif},\ and\ \citenamefont
  {Esslinger}}]{jotzu.nature.515.237.2014}%
  \BibitemOpen
  \bibfield  {author} {\bibinfo {author} {\bibfnamefont {G.}~\bibnamefont
  {Jotzu}}, \bibinfo {author} {\bibfnamefont {M.}~\bibnamefont {Messer}},
  \bibinfo {author} {\bibfnamefont {R.}~\bibnamefont {Desbuquois}}, \bibinfo
  {author} {\bibfnamefont {M.}~\bibnamefont {Lebrat}}, \bibinfo {author}
  {\bibfnamefont {T.}~\bibnamefont {Uehlinger}}, \bibinfo {author}
  {\bibfnamefont {D.}~\bibnamefont {Greif}}, \ and\ \bibinfo {author}
  {\bibfnamefont {T.}~\bibnamefont {Esslinger}},\ }\href {\doibase
  10.1038/nature13915} {\bibfield  {journal} {\bibinfo  {journal} {Nature}\
  }\textbf {\bibinfo {volume} {515}},\ \bibinfo {pages} {237} (\bibinfo {year}
  {2014})}\BibitemShut {NoStop}%
\bibitem [{\citenamefont {Song}\ \emph {et~al.}(2015)\citenamefont {Song},
  \citenamefont {Paltoglou}, \citenamefont {Liu}, \citenamefont {Zhu},
  \citenamefont {Gallardo}, \citenamefont {Tang}, \citenamefont {Xu},
  \citenamefont {Ablowitz}, \citenamefont {Efremidis},\ and\ \citenamefont
  {Chen}}]{song.nc.6.6272.2015}%
  \BibitemOpen
  \bibfield  {author} {\bibinfo {author} {\bibfnamefont {D.}~\bibnamefont
  {Song}}, \bibinfo {author} {\bibfnamefont {V.}~\bibnamefont {Paltoglou}},
  \bibinfo {author} {\bibfnamefont {S.}~\bibnamefont {Liu}}, \bibinfo {author}
  {\bibfnamefont {Y.}~\bibnamefont {Zhu}}, \bibinfo {author} {\bibfnamefont
  {D.}~\bibnamefont {Gallardo}}, \bibinfo {author} {\bibfnamefont
  {L.}~\bibnamefont {Tang}}, \bibinfo {author} {\bibfnamefont {J.}~\bibnamefont
  {Xu}}, \bibinfo {author} {\bibfnamefont {M.}~\bibnamefont {Ablowitz}},
  \bibinfo {author} {\bibfnamefont {N.~K.}\ \bibnamefont {Efremidis}}, \ and\
  \bibinfo {author} {\bibfnamefont {Z.}~\bibnamefont {Chen}},\ }\href {\doibase
  10.1038/ncomms7272} {\bibfield  {journal} {\bibinfo  {journal} {Nat.
  Commun.}\ }\textbf {\bibinfo {volume} {6}},\ \bibinfo {pages} {6272}
  (\bibinfo {year} {2015})}\BibitemShut {NoStop}%
\bibitem [{\citenamefont {Kane}\ and\ \citenamefont
  {Mele}(2005)}]{kane.prl.95.226801.2005}%
  \BibitemOpen
  \bibfield  {author} {\bibinfo {author} {\bibfnamefont {C.~L.}\ \bibnamefont
  {Kane}}\ and\ \bibinfo {author} {\bibfnamefont {E.~J.}\ \bibnamefont
  {Mele}},\ }\href {\doibase 10.1103/PhysRevLett.95.226801} {\bibfield
  {journal} {\bibinfo  {journal} {Phys. Rev. Lett.}\ }\textbf {\bibinfo
  {volume} {95}},\ \bibinfo {pages} {226801} (\bibinfo {year}
  {2005})}\BibitemShut {NoStop}%
\bibitem [{\citenamefont {Hsieh}\ \emph {et~al.}(2008)\citenamefont {Hsieh},
  \citenamefont {Qian}, \citenamefont {Wray}, \citenamefont {Xia},
  \citenamefont {Hor}, \citenamefont {Cava},\ and\ \citenamefont
  {Hasan}}]{hsieh.nature.452.970.2008}%
  \BibitemOpen
  \bibfield  {author} {\bibinfo {author} {\bibfnamefont {D.}~\bibnamefont
  {Hsieh}}, \bibinfo {author} {\bibfnamefont {D.}~\bibnamefont {Qian}},
  \bibinfo {author} {\bibfnamefont {L.}~\bibnamefont {Wray}}, \bibinfo {author}
  {\bibfnamefont {Y.}~\bibnamefont {Xia}}, \bibinfo {author} {\bibfnamefont
  {Y.~S.}\ \bibnamefont {Hor}}, \bibinfo {author} {\bibfnamefont {R.~J.}\
  \bibnamefont {Cava}}, \ and\ \bibinfo {author} {\bibfnamefont {M.~Z.}\
  \bibnamefont {Hasan}},\ }\href {\doibase 10.1038/nature06843} {\bibfield
  {journal} {\bibinfo  {journal} {Nature}\ }\textbf {\bibinfo {volume} {452}},\
  \bibinfo {pages} {970} (\bibinfo {year} {2008})}\BibitemShut {NoStop}%
\bibitem [{\citenamefont {Hasan}\ and\ \citenamefont
  {Kane}(2010)}]{hasan.rmp.82.3045.2010}%
  \BibitemOpen
  \bibfield  {author} {\bibinfo {author} {\bibfnamefont {M.~Z.}\ \bibnamefont
  {Hasan}}\ and\ \bibinfo {author} {\bibfnamefont {C.~L.}\ \bibnamefont
  {Kane}},\ }\href {\doibase 10.1103/RevModPhys.82.3045} {\bibfield  {journal}
  {\bibinfo  {journal} {Rev. Mod. Phys.}\ }\textbf {\bibinfo {volume} {82}},\
  \bibinfo {pages} {3045} (\bibinfo {year} {2010})}\BibitemShut {NoStop}%
\bibitem [{\citenamefont {Qi}\ and\ \citenamefont
  {Zhang}(2011)}]{qi.rmp.83.1057.2011}%
  \BibitemOpen
  \bibfield  {author} {\bibinfo {author} {\bibfnamefont {X.-L.}\ \bibnamefont
  {Qi}}\ and\ \bibinfo {author} {\bibfnamefont {S.-C.}\ \bibnamefont {Zhang}},\
  }\href {\doibase 10.1103/RevModPhys.83.1057} {\bibfield  {journal} {\bibinfo
  {journal} {Rev. Mod. Phys.}\ }\textbf {\bibinfo {volume} {83}},\ \bibinfo
  {pages} {1057} (\bibinfo {year} {2011})}\BibitemShut {NoStop}%
\bibitem [{\citenamefont {Liu}\ \emph {et~al.}(2011)\citenamefont {Liu},
  \citenamefont {Yin}, \citenamefont {Ulin-Avila}, \citenamefont {Geng},
  \citenamefont {Zentgraf}, \citenamefont {Ju}, \citenamefont {Wang},\ and\
  \citenamefont {Zhang}}]{liu.nature.474.64.2011}%
  \BibitemOpen
  \bibfield  {author} {\bibinfo {author} {\bibfnamefont {M.}~\bibnamefont
  {Liu}}, \bibinfo {author} {\bibfnamefont {X.}~\bibnamefont {Yin}}, \bibinfo
  {author} {\bibfnamefont {E.}~\bibnamefont {Ulin-Avila}}, \bibinfo {author}
  {\bibfnamefont {B.}~\bibnamefont {Geng}}, \bibinfo {author} {\bibfnamefont
  {T.}~\bibnamefont {Zentgraf}}, \bibinfo {author} {\bibfnamefont
  {L.}~\bibnamefont {Ju}}, \bibinfo {author} {\bibfnamefont {F.}~\bibnamefont
  {Wang}}, \ and\ \bibinfo {author} {\bibfnamefont {X.}~\bibnamefont {Zhang}},\
  }\href {\doibase 10.1038/nature10067} {\bibfield  {journal} {\bibinfo
  {journal} {Nature}\ }\textbf {\bibinfo {volume} {474}},\ \bibinfo {pages}
  {64} (\bibinfo {year} {2011})}\BibitemShut {NoStop}%
\bibitem [{\citenamefont {Liang}\ and\ \citenamefont
  {Chong}(2013)}]{liang.prl.110.203904.2013}%
  \BibitemOpen
  \bibfield  {author} {\bibinfo {author} {\bibfnamefont {G.~Q.}\ \bibnamefont
  {Liang}}\ and\ \bibinfo {author} {\bibfnamefont {Y.~D.}\ \bibnamefont
  {Chong}},\ }\href {\doibase 10.1103/PhysRevLett.110.203904} {\bibfield
  {journal} {\bibinfo  {journal} {Phys. Rev. Lett.}\ }\textbf {\bibinfo
  {volume} {110}},\ \bibinfo {pages} {203904} (\bibinfo {year}
  {2013})}\BibitemShut {NoStop}%
\bibitem [{\citenamefont {Zhang}\ \emph {et~al.}(2015)\citenamefont {Zhang},
  \citenamefont {Wu}, \citenamefont {Beli\'c}, \citenamefont {Zheng},
  \citenamefont {Wang}, \citenamefont {Xiao},\ and\ \citenamefont
  {Zhang}}]{zhang.lpr.9.1.2015}%
  \BibitemOpen
  \bibfield  {author} {\bibinfo {author} {\bibfnamefont {Y.~Q.}\ \bibnamefont
  {Zhang}}, \bibinfo {author} {\bibfnamefont {Z.~K.}\ \bibnamefont {Wu}},
  \bibinfo {author} {\bibfnamefont {M.~R.}\ \bibnamefont {Beli\'c}}, \bibinfo
  {author} {\bibfnamefont {H.~B.}\ \bibnamefont {Zheng}}, \bibinfo {author}
  {\bibfnamefont {Z.~G.}\ \bibnamefont {Wang}}, \bibinfo {author}
  {\bibfnamefont {M.}~\bibnamefont {Xiao}}, \ and\ \bibinfo {author}
  {\bibfnamefont {Y.~P.}\ \bibnamefont {Zhang}},\ }\href {\doibase
  10.1002/lpor.201400428} {\bibfield  {journal} {\bibinfo  {journal} {Laser
  Photonics Rev.}\ }\textbf {\bibinfo {volume} {9}},\ \bibinfo {pages} {331}
  (\bibinfo {year} {2015})}\BibitemShut {NoStop}%
\bibitem [{\citenamefont {Singha}\ \emph {et~al.}(2011)\citenamefont {Singha},
  \citenamefont {Gibertini}, \citenamefont {Karmakar}, \citenamefont {Yuan},
  \citenamefont {Polini}, \citenamefont {Vignale}, \citenamefont {Katsnelson},
  \citenamefont {Pinczuk}, \citenamefont {Pfeiffer}, \citenamefont {West},\
  and\ \citenamefont {Pellegrini}}]{singha.science.332.1176.2011}%
  \BibitemOpen
  \bibfield  {author} {\bibinfo {author} {\bibfnamefont {A.}~\bibnamefont
  {Singha}}, \bibinfo {author} {\bibfnamefont {M.}~\bibnamefont {Gibertini}},
  \bibinfo {author} {\bibfnamefont {B.}~\bibnamefont {Karmakar}}, \bibinfo
  {author} {\bibfnamefont {S.}~\bibnamefont {Yuan}}, \bibinfo {author}
  {\bibfnamefont {M.}~\bibnamefont {Polini}}, \bibinfo {author} {\bibfnamefont
  {G.}~\bibnamefont {Vignale}}, \bibinfo {author} {\bibfnamefont {M.~I.}\
  \bibnamefont {Katsnelson}}, \bibinfo {author} {\bibfnamefont
  {A.}~\bibnamefont {Pinczuk}}, \bibinfo {author} {\bibfnamefont {L.~N.}\
  \bibnamefont {Pfeiffer}}, \bibinfo {author} {\bibfnamefont {K.~W.}\
  \bibnamefont {West}}, \ and\ \bibinfo {author} {\bibfnamefont
  {V.}~\bibnamefont {Pellegrini}},\ }\href {\doibase 10.1126/science.1204333}
  {\bibfield  {journal} {\bibinfo  {journal} {Science}\ }\textbf {\bibinfo
  {volume} {332}},\ \bibinfo {pages} {1176} (\bibinfo {year}
  {2011})}\BibitemShut {NoStop}%
\bibitem [{\citenamefont {Zhao}\ and\ \citenamefont
  {Paramekanti}(2006)}]{zhao.prl.97.230404.2006}%
  \BibitemOpen
  \bibfield  {author} {\bibinfo {author} {\bibfnamefont {E.}~\bibnamefont
  {Zhao}}\ and\ \bibinfo {author} {\bibfnamefont {A.}~\bibnamefont
  {Paramekanti}},\ }\href {\doibase 10.1103/PhysRevLett.97.230404} {\bibfield
  {journal} {\bibinfo  {journal} {Phys. Rev. Lett.}\ }\textbf {\bibinfo
  {volume} {97}},\ \bibinfo {pages} {230404} (\bibinfo {year}
  {2006})}\BibitemShut {NoStop}%
\bibitem [{\citenamefont {Wu}\ and\ \citenamefont
  {Das~Sarma}(2008)}]{wu.prb.77.235107.2008}%
  \BibitemOpen
  \bibfield  {author} {\bibinfo {author} {\bibfnamefont {C.}~\bibnamefont
  {Wu}}\ and\ \bibinfo {author} {\bibfnamefont {S.}~\bibnamefont {Das~Sarma}},\
  }\href {\doibase 10.1103/PhysRevB.77.235107} {\bibfield  {journal} {\bibinfo
  {journal} {Phys. Rev. B}\ }\textbf {\bibinfo {volume} {77}},\ \bibinfo
  {pages} {235107} (\bibinfo {year} {2008})}\BibitemShut {NoStop}%
\bibitem [{\citenamefont {Artoni}\ and\ \citenamefont
  {La~Rocca}(2006)}]{artoni.prl.96.073905.2006}%
  \BibitemOpen
  \bibfield  {author} {\bibinfo {author} {\bibfnamefont {M.}~\bibnamefont
  {Artoni}}\ and\ \bibinfo {author} {\bibfnamefont {G.~C.}\ \bibnamefont
  {La~Rocca}},\ }\href {\doibase 10.1103/PhysRevLett.96.073905} {\bibfield
  {journal} {\bibinfo  {journal} {Phys. Rev. Lett.}\ }\textbf {\bibinfo
  {volume} {96}},\ \bibinfo {pages} {073905} (\bibinfo {year}
  {2006})}\BibitemShut {NoStop}%
\bibitem [{\citenamefont {Zhang}\ \emph
  {et~al.}(2012{\natexlab{a}})\citenamefont {Zhang}, \citenamefont {Yao},
  \citenamefont {Yuan}, \citenamefont {Li}, \citenamefont {Yuan}, \citenamefont
  {Feng}, \citenamefont {Jia},\ and\ \citenamefont
  {Zhang}}]{zhang.ieee.4.2057.2012}%
  \BibitemOpen
  \bibfield  {author} {\bibinfo {author} {\bibfnamefont {Y.~Q.}\ \bibnamefont
  {Zhang}}, \bibinfo {author} {\bibfnamefont {X.}~\bibnamefont {Yao}}, \bibinfo
  {author} {\bibfnamefont {C.~Z.}\ \bibnamefont {Yuan}}, \bibinfo {author}
  {\bibfnamefont {P.~Y.}\ \bibnamefont {Li}}, \bibinfo {author} {\bibfnamefont
  {J.~M.}\ \bibnamefont {Yuan}}, \bibinfo {author} {\bibfnamefont {W.~K.}\
  \bibnamefont {Feng}}, \bibinfo {author} {\bibfnamefont {S.~Q.}\ \bibnamefont
  {Jia}}, \ and\ \bibinfo {author} {\bibfnamefont {Y.~P.}\ \bibnamefont
  {Zhang}},\ }\href {\doibase 10.1109/JPHOT.2012.2225609} {\bibfield  {journal}
  {\bibinfo  {journal} {IEEE Photon. J.}\ }\textbf {\bibinfo {volume} {4}},\
  \bibinfo {pages} {2057} (\bibinfo {year} {2012}{\natexlab{a}})}\BibitemShut
  {NoStop}%
\bibitem [{\citenamefont {Wang}\ \emph {et~al.}(2013)\citenamefont {Wang},
  \citenamefont {Zhou}, \citenamefont {Guo}, \citenamefont {Zhang},
  \citenamefont {Evers},\ and\ \citenamefont {Zhu}}]{wang.prl.110.093901.2013}%
  \BibitemOpen
  \bibfield  {author} {\bibinfo {author} {\bibfnamefont {D.-W.}\ \bibnamefont
  {Wang}}, \bibinfo {author} {\bibfnamefont {H.-T.}\ \bibnamefont {Zhou}},
  \bibinfo {author} {\bibfnamefont {M.-J.}\ \bibnamefont {Guo}}, \bibinfo
  {author} {\bibfnamefont {J.-X.}\ \bibnamefont {Zhang}}, \bibinfo {author}
  {\bibfnamefont {J.}~\bibnamefont {Evers}}, \ and\ \bibinfo {author}
  {\bibfnamefont {S.-Y.}\ \bibnamefont {Zhu}},\ }\href {\doibase
  10.1103/PhysRevLett.110.093901} {\bibfield  {journal} {\bibinfo  {journal}
  {Phys. Rev. Lett.}\ }\textbf {\bibinfo {volume} {110}},\ \bibinfo {pages}
  {093901} (\bibinfo {year} {2013})}\BibitemShut {NoStop}%
\bibitem [{\citenamefont {Zhang}\ \emph {et~al.}(2014)\citenamefont {Zhang},
  \citenamefont {Lan}, \citenamefont {Li}, \citenamefont {Zheng}, \citenamefont
  {Lei}, \citenamefont {Wang}, \citenamefont {Metlo},\ and\ \citenamefont
  {Zhang}}]{zhang.jpcc.118.14521.2014}%
  \BibitemOpen
  \bibfield  {author} {\bibinfo {author} {\bibfnamefont {D.}~\bibnamefont
  {Zhang}}, \bibinfo {author} {\bibfnamefont {H.~Y.}\ \bibnamefont {Lan}},
  \bibinfo {author} {\bibfnamefont {C.~B.}\ \bibnamefont {Li}}, \bibinfo
  {author} {\bibfnamefont {H.~B.}\ \bibnamefont {Zheng}}, \bibinfo {author}
  {\bibfnamefont {C.~J.}\ \bibnamefont {Lei}}, \bibinfo {author} {\bibfnamefont
  {R.~M.}\ \bibnamefont {Wang}}, \bibinfo {author} {\bibfnamefont
  {I.}~\bibnamefont {Metlo}}, \ and\ \bibinfo {author} {\bibfnamefont {Y.~P.}\
  \bibnamefont {Zhang}},\ }\href {\doibase 10.1021/jp502343g} {\bibfield
  {journal} {\bibinfo  {journal} {J. Phys. Chem. C}\ }\textbf {\bibinfo
  {volume} {118}},\ \bibinfo {pages} {14521} (\bibinfo {year}
  {2014})}\BibitemShut {NoStop}%
\bibitem [{\citenamefont {Lan}\ \emph {et~al.}(2015)\citenamefont {Lan},
  \citenamefont {Li}, \citenamefont {Lei}, \citenamefont {Zheng}, \citenamefont
  {Wang}, \citenamefont {Xiao},\ and\ \citenamefont
  {Zhang}}]{lan.lpl.12.015404.2015}%
  \BibitemOpen
  \bibfield  {author} {\bibinfo {author} {\bibfnamefont {H.~Y.}\ \bibnamefont
  {Lan}}, \bibinfo {author} {\bibfnamefont {C.~B.}\ \bibnamefont {Li}},
  \bibinfo {author} {\bibfnamefont {C.~J.}\ \bibnamefont {Lei}}, \bibinfo
  {author} {\bibfnamefont {H.~B.}\ \bibnamefont {Zheng}}, \bibinfo {author}
  {\bibfnamefont {R.~M.}\ \bibnamefont {Wang}}, \bibinfo {author}
  {\bibfnamefont {M.}~\bibnamefont {Xiao}}, \ and\ \bibinfo {author}
  {\bibfnamefont {Y.~P.}\ \bibnamefont {Zhang}},\ }\href
  {http://stacks.iop.org/1612-202X/12/i=1/a=015404} {\bibfield  {journal}
  {\bibinfo  {journal} {Laser Phys. Lett.}\ }\textbf {\bibinfo {volume} {12}},\
  \bibinfo {pages} {015404} (\bibinfo {year} {2015})}\BibitemShut {NoStop}%
\bibitem [{\citenamefont {Li}\ \emph {et~al.}(2015)\citenamefont {Li},
  \citenamefont {Wang}, \citenamefont {Yang}, \citenamefont {Jiang},
  \citenamefont {Maitlo}, \citenamefont {Ahmed}, \citenamefont {Xiao},\ and\
  \citenamefont {Zhang}}]{li.rsc.2015}%
  \BibitemOpen
  \bibfield  {author} {\bibinfo {author} {\bibfnamefont {C.~B.}\ \bibnamefont
  {Li}}, \bibinfo {author} {\bibfnamefont {L.~L.}\ \bibnamefont {Wang}},
  \bibinfo {author} {\bibfnamefont {C.}~\bibnamefont {Yang}}, \bibinfo {author}
  {\bibfnamefont {T.}~\bibnamefont {Jiang}}, \bibinfo {author} {\bibfnamefont
  {I.}~\bibnamefont {Maitlo}}, \bibinfo {author} {\bibfnamefont
  {I.}~\bibnamefont {Ahmed}}, \bibinfo {author} {\bibfnamefont
  {M.}~\bibnamefont {Xiao}}, \ and\ \bibinfo {author} {\bibfnamefont {Y.~P.}\
  \bibnamefont {Zhang}},\ }\href {\doibase 10.1039/C5RA05607A} {\bibfield
  {journal} {\bibinfo  {journal} {RSC Adv.}\ }\textbf {\bibinfo {volume} {5}},\
  \bibinfo {pages} {39449} (\bibinfo {year} {2015})}\BibitemShut {NoStop}%
\bibitem [{\citenamefont {Phillips}\ \emph {et~al.}(2001)\citenamefont
  {Phillips}, \citenamefont {Fleischhauer}, \citenamefont {Mair}, \citenamefont
  {Walsworth},\ and\ \citenamefont {Lukin}}]{philips.prl.86.783.2001}%
  \BibitemOpen
  \bibfield  {author} {\bibinfo {author} {\bibfnamefont {D.~F.}\ \bibnamefont
  {Phillips}}, \bibinfo {author} {\bibfnamefont {A.}~\bibnamefont
  {Fleischhauer}}, \bibinfo {author} {\bibfnamefont {A.}~\bibnamefont {Mair}},
  \bibinfo {author} {\bibfnamefont {R.~L.}\ \bibnamefont {Walsworth}}, \ and\
  \bibinfo {author} {\bibfnamefont {M.~D.}\ \bibnamefont {Lukin}},\ }\href
  {\doibase 10.1103/PhysRevLett.86.783} {\bibfield  {journal} {\bibinfo
  {journal} {Phys. Rev. Lett.}\ }\textbf {\bibinfo {volume} {86}},\ \bibinfo
  {pages} {783} (\bibinfo {year} {2001})}\BibitemShut {NoStop}%
\bibitem [{\citenamefont {Hang}\ \emph {et~al.}(2013)\citenamefont {Hang},
  \citenamefont {Huang},\ and\ \citenamefont
  {Konotop}}]{hang.prl.110.083604.2013}%
  \BibitemOpen
  \bibfield  {author} {\bibinfo {author} {\bibfnamefont {C.}~\bibnamefont
  {Hang}}, \bibinfo {author} {\bibfnamefont {G.}~\bibnamefont {Huang}}, \ and\
  \bibinfo {author} {\bibfnamefont {V.~V.}\ \bibnamefont {Konotop}},\ }\href
  {\doibase 10.1103/PhysRevLett.110.083604} {\bibfield  {journal} {\bibinfo
  {journal} {Phys. Rev. Lett.}\ }\textbf {\bibinfo {volume} {110}},\ \bibinfo
  {pages} {083604} (\bibinfo {year} {2013})}\BibitemShut {NoStop}%
\bibitem [{\citenamefont {Wu}\ \emph {et~al.}(2013)\citenamefont {Wu},
  \citenamefont {Horsley}, \citenamefont {Artoni},\ and\ \citenamefont
  {La~Rocca}}]{wu.light.2.e54.2013}%
  \BibitemOpen
  \bibfield  {author} {\bibinfo {author} {\bibfnamefont {J.-H.}\ \bibnamefont
  {Wu}}, \bibinfo {author} {\bibfnamefont {S.~A.~R.}\ \bibnamefont {Horsley}},
  \bibinfo {author} {\bibfnamefont {M.}~\bibnamefont {Artoni}}, \ and\ \bibinfo
  {author} {\bibfnamefont {G.~C.}\ \bibnamefont {La~Rocca}},\ }\href {\doibase
  10.1038/lsa.2013.10} {\bibfield  {journal} {\bibinfo  {journal} {Light: Sci.
  \& Appl.}\ }\textbf {\bibinfo {volume} {2}},\ \bibinfo {pages} {e54}
  (\bibinfo {year} {2013})}\BibitemShut {NoStop}%
\bibitem [{\citenamefont {Horsley}\ \emph {et~al.}(2013)\citenamefont
  {Horsley}, \citenamefont {Wu}, \citenamefont {Artoni},\ and\ \citenamefont
  {La~Rocca}}]{horesley.prl.110.223602.2013}%
  \BibitemOpen
  \bibfield  {author} {\bibinfo {author} {\bibfnamefont {S.~A.~R.}\
  \bibnamefont {Horsley}}, \bibinfo {author} {\bibfnamefont {J.-H.}\
  \bibnamefont {Wu}}, \bibinfo {author} {\bibfnamefont {M.}~\bibnamefont
  {Artoni}}, \ and\ \bibinfo {author} {\bibfnamefont {G.~C.}\ \bibnamefont
  {La~Rocca}},\ }\href {\doibase 10.1103/PhysRevLett.110.223602} {\bibfield
  {journal} {\bibinfo  {journal} {Phys. Rev. Lett.}\ }\textbf {\bibinfo
  {volume} {110}},\ \bibinfo {pages} {223602} (\bibinfo {year}
  {2013})}\BibitemShut {NoStop}%
\bibitem [{\citenamefont {Yan}\ \emph {et~al.}(2001)\citenamefont {Yan},
  \citenamefont {Rickey},\ and\ \citenamefont {Zhu}}]{yan.pra.64.013412.2001}%
  \BibitemOpen
  \bibfield  {author} {\bibinfo {author} {\bibfnamefont {M.}~\bibnamefont
  {Yan}}, \bibinfo {author} {\bibfnamefont {E.~G.}\ \bibnamefont {Rickey}}, \
  and\ \bibinfo {author} {\bibfnamefont {Y.}~\bibnamefont {Zhu}},\ }\href
  {\doibase 10.1103/PhysRevA.64.013412} {\bibfield  {journal} {\bibinfo
  {journal} {Phys. Rev. A}\ }\textbf {\bibinfo {volume} {64}},\ \bibinfo
  {pages} {013412} (\bibinfo {year} {2001})}\BibitemShut {NoStop}%
\bibitem [{\citenamefont {Sun}\ \emph {et~al.}(2004)\citenamefont {Sun},
  \citenamefont {Zuo}, \citenamefont {Mi}, \citenamefont {Yu}, \citenamefont
  {Jiang}, \citenamefont {Wang}, \citenamefont {Wu},\ and\ \citenamefont
  {Fu}}]{sun.pra.70.053820.2004}%
  \BibitemOpen
  \bibfield  {author} {\bibinfo {author} {\bibfnamefont {J.}~\bibnamefont
  {Sun}}, \bibinfo {author} {\bibfnamefont {Z.}~\bibnamefont {Zuo}}, \bibinfo
  {author} {\bibfnamefont {X.}~\bibnamefont {Mi}}, \bibinfo {author}
  {\bibfnamefont {Z.}~\bibnamefont {Yu}}, \bibinfo {author} {\bibfnamefont
  {Q.}~\bibnamefont {Jiang}}, \bibinfo {author} {\bibfnamefont
  {Y.}~\bibnamefont {Wang}}, \bibinfo {author} {\bibfnamefont {L.-A.}\
  \bibnamefont {Wu}}, \ and\ \bibinfo {author} {\bibfnamefont {P.}~\bibnamefont
  {Fu}},\ }\href {\doibase 10.1103/PhysRevA.70.053820} {\bibfield  {journal}
  {\bibinfo  {journal} {Phys. Rev. A}\ }\textbf {\bibinfo {volume} {70}},\
  \bibinfo {pages} {053820} (\bibinfo {year} {2004})}\BibitemShut {NoStop}%
\bibitem [{\citenamefont {Wang}\ \emph {et~al.}(2004)\citenamefont {Wang},
  \citenamefont {Kong}, \citenamefont {Tu}, \citenamefont {Jiang},
  \citenamefont {Li}, \citenamefont {Xiong}, \citenamefont {Zhu},\ and\
  \citenamefont {Zhan}}]{wang.pla.328.437.2004}%
  \BibitemOpen
  \bibfield  {author} {\bibinfo {author} {\bibfnamefont {J.}~\bibnamefont
  {Wang}}, \bibinfo {author} {\bibfnamefont {L.}~\bibnamefont {Kong}}, \bibinfo
  {author} {\bibfnamefont {X.}~\bibnamefont {Tu}}, \bibinfo {author}
  {\bibfnamefont {K.}~\bibnamefont {Jiang}}, \bibinfo {author} {\bibfnamefont
  {K.}~\bibnamefont {Li}}, \bibinfo {author} {\bibfnamefont {H.}~\bibnamefont
  {Xiong}}, \bibinfo {author} {\bibfnamefont {Y.}~\bibnamefont {Zhu}}, \ and\
  \bibinfo {author} {\bibfnamefont {M.}~\bibnamefont {Zhan}},\ }\href {\doibase
  http://dx.doi.org/10.1016/j.physleta.2004.06.049} {\bibfield  {journal}
  {\bibinfo  {journal} {Phys. Lett. A}\ }\textbf {\bibinfo {volume} {328}},\
  \bibinfo {pages} {437} (\bibinfo {year} {2004})}\BibitemShut {NoStop}%
\bibitem [{\citenamefont {Niu}\ \emph {et~al.}(2011)\citenamefont {Niu},
  \citenamefont {Pei}, \citenamefont {Lu}, \citenamefont {Wang}, \citenamefont
  {Wu},\ and\ \citenamefont {Fu}}]{niu.pra.84.033853.2011}%
  \BibitemOpen
  \bibfield  {author} {\bibinfo {author} {\bibfnamefont {J.}~\bibnamefont
  {Niu}}, \bibinfo {author} {\bibfnamefont {L.}~\bibnamefont {Pei}}, \bibinfo
  {author} {\bibfnamefont {X.}~\bibnamefont {Lu}}, \bibinfo {author}
  {\bibfnamefont {R.}~\bibnamefont {Wang}}, \bibinfo {author} {\bibfnamefont
  {L.-A.}\ \bibnamefont {Wu}}, \ and\ \bibinfo {author} {\bibfnamefont
  {P.}~\bibnamefont {Fu}},\ }\href {\doibase 10.1103/PhysRevA.84.033853}
  {\bibfield  {journal} {\bibinfo  {journal} {Phys. Rev. A}\ }\textbf {\bibinfo
  {volume} {84}},\ \bibinfo {pages} {033853} (\bibinfo {year}
  {2011})}\BibitemShut {NoStop}%
\bibitem [{\citenamefont {Michinel}\ \emph {et~al.}(2006)\citenamefont
  {Michinel}, \citenamefont {Paz-Alonso},\ and\ \citenamefont
  {P{\'e}rez-Garc{\'i}a}}]{michinel.prl.96.023903.2006}%
  \BibitemOpen
  \bibfield  {author} {\bibinfo {author} {\bibfnamefont {H.}~\bibnamefont
  {Michinel}}, \bibinfo {author} {\bibfnamefont {M.~J.}\ \bibnamefont
  {Paz-Alonso}}, \ and\ \bibinfo {author} {\bibfnamefont {V.~M.}\ \bibnamefont
  {P{\'e}rez-Garc{\'i}a}},\ }\href {\doibase 10.1103/PhysRevLett.96.023903}
  {\bibfield  {journal} {\bibinfo  {journal} {Phys. Rev. Lett.}\ }\textbf
  {\bibinfo {volume} {96}},\ \bibinfo {pages} {023903} (\bibinfo {year}
  {2006})}\BibitemShut {NoStop}%
\bibitem [{\citenamefont {Zhang}\ \emph
  {et~al.}(2012{\natexlab{b}})\citenamefont {Zhang}, \citenamefont {Wu},
  \citenamefont {Yuan}, \citenamefont {Yao}, \citenamefont {Lu}, \citenamefont
  {Beli\'{c}},\ and\ \citenamefont {Zhang}}]{zhang.ol.37.4507.2012}%
  \BibitemOpen
  \bibfield  {author} {\bibinfo {author} {\bibfnamefont {Y.~Q.}\ \bibnamefont
  {Zhang}}, \bibinfo {author} {\bibfnamefont {Z.~K.}\ \bibnamefont {Wu}},
  \bibinfo {author} {\bibfnamefont {C.~Z.}\ \bibnamefont {Yuan}}, \bibinfo
  {author} {\bibfnamefont {X.}~\bibnamefont {Yao}}, \bibinfo {author}
  {\bibfnamefont {K.~Q.}\ \bibnamefont {Lu}}, \bibinfo {author} {\bibfnamefont
  {M.}~\bibnamefont {Beli\'{c}}}, \ and\ \bibinfo {author} {\bibfnamefont
  {Y.~P.}\ \bibnamefont {Zhang}},\ }\href {\doibase 10.1364/OL.37.004507}
  {\bibfield  {journal} {\bibinfo  {journal} {Opt. Lett.}\ }\textbf {\bibinfo
  {volume} {37}},\ \bibinfo {pages} {4507} (\bibinfo {year}
  {2012}{\natexlab{b}})}\BibitemShut {NoStop}%
\bibitem [{\citenamefont {Zhang}\ \emph {et~al.}(2013)\citenamefont {Zhang},
  \citenamefont {Beli\'{c}}, \citenamefont {Wu}, \citenamefont {Yuan},
  \citenamefont {Wang}, \citenamefont {Lu},\ and\ \citenamefont
  {Zhang}}]{zhang.pra.88.013847.2013}%
  \BibitemOpen
  \bibfield  {author} {\bibinfo {author} {\bibfnamefont {Y.~Q.}\ \bibnamefont
  {Zhang}}, \bibinfo {author} {\bibfnamefont {M.}~\bibnamefont {Beli\'{c}}},
  \bibinfo {author} {\bibfnamefont {Z.~K.}\ \bibnamefont {Wu}}, \bibinfo
  {author} {\bibfnamefont {C.~Z.}\ \bibnamefont {Yuan}}, \bibinfo {author}
  {\bibfnamefont {R.~M.}\ \bibnamefont {Wang}}, \bibinfo {author}
  {\bibfnamefont {K.~Q.}\ \bibnamefont {Lu}}, \ and\ \bibinfo {author}
  {\bibfnamefont {Y.~P.}\ \bibnamefont {Zhang}},\ }\href {\doibase
  10.1103/PhysRevA.88.013847} {\bibfield  {journal} {\bibinfo  {journal} {Phys.
  Rev. A}\ }\textbf {\bibinfo {volume} {88}},\ \bibinfo {pages} {013847}
  (\bibinfo {year} {2013})}\BibitemShut {NoStop}%
\bibitem [{\citenamefont {Paredes}\ \emph {et~al.}(2014)\citenamefont
  {Paredes}, \citenamefont {Feijoo},\ and\ \citenamefont
  {Michinel}}]{paredes.prl.112.173901.2014}%
  \BibitemOpen
  \bibfield  {author} {\bibinfo {author} {\bibfnamefont {A.}~\bibnamefont
  {Paredes}}, \bibinfo {author} {\bibfnamefont {D.}~\bibnamefont {Feijoo}}, \
  and\ \bibinfo {author} {\bibfnamefont {H.}~\bibnamefont {Michinel}},\ }\href
  {\doibase 10.1103/PhysRevLett.112.173901} {\bibfield  {journal} {\bibinfo
  {journal} {Phys. Rev. Lett.}\ }\textbf {\bibinfo {volume} {112}},\ \bibinfo
  {pages} {173901} (\bibinfo {year} {2014})}\BibitemShut {NoStop}%
\bibitem [{\citenamefont {Harris}(1997)}]{harris.pt.50.36.1997}%
  \BibitemOpen
  \bibfield  {author} {\bibinfo {author} {\bibfnamefont {S.~E.}\ \bibnamefont
  {Harris}},\ }\href {\doibase 10.1063/1.881806} {\bibfield  {journal}
  {\bibinfo  {journal} {Phys. Today}\ }\textbf {\bibinfo {volume} {50(7)}},\
  \bibinfo {pages} {36} (\bibinfo {year} {1997})}\BibitemShut {NoStop}%
\bibitem [{\citenamefont {Fleischhauer}\ \emph {et~al.}(2005)\citenamefont
  {Fleischhauer}, \citenamefont {Imamoglu},\ and\ \citenamefont
  {Marangos}}]{fleischhauer.rmp.77.633.2005}%
  \BibitemOpen
  \bibfield  {author} {\bibinfo {author} {\bibfnamefont {M.}~\bibnamefont
  {Fleischhauer}}, \bibinfo {author} {\bibfnamefont {A.}~\bibnamefont
  {Imamoglu}}, \ and\ \bibinfo {author} {\bibfnamefont {J.~P.}\ \bibnamefont
  {Marangos}},\ }\href {\doibase 10.1103/RevModPhys.77.633} {\bibfield
  {journal} {\bibinfo  {journal} {Rev. Mod. Phys.}\ }\textbf {\bibinfo {volume}
  {77}},\ \bibinfo {pages} {633} (\bibinfo {year} {2005})}\BibitemShut
  {NoStop}%
\bibitem [{\citenamefont {Ablowitz}\ and\ \citenamefont
  {Zhu}(2012)}]{ablowitz.jam.72.240.2012}%
  \BibitemOpen
  \bibfield  {author} {\bibinfo {author} {\bibfnamefont {M.}~\bibnamefont
  {Ablowitz}}\ and\ \bibinfo {author} {\bibfnamefont {Y.}~\bibnamefont {Zhu}},\
  }\href {\doibase 10.1137/11082662X} {\bibfield  {journal} {\bibinfo
  {journal} {SIAM J. Appl. Math.}\ }\textbf {\bibinfo {volume} {72}},\ \bibinfo
  {pages} {240} (\bibinfo {year} {2012})}\BibitemShut {NoStop}%
\bibitem [{\citenamefont {Zhang}\ and\ \citenamefont
  {Xiao}(2009)}]{zhang.book.2009}%
  \BibitemOpen
  \bibfield  {author} {\bibinfo {author} {\bibfnamefont {Y.}~\bibnamefont
  {Zhang}}\ and\ \bibinfo {author} {\bibfnamefont {M.}~\bibnamefont {Xiao}},\
  }\href@noop {} {\emph {\bibinfo {title} {Multi-wave mixing processes: from
  ultrafast polarization beats to electromagnetically induced transparency}}}\
  (\bibinfo  {publisher} {Springer},\ \bibinfo {address} {New York},\ \bibinfo
  {year} {2009})\BibitemShut {NoStop}%
\bibitem [{\citenamefont {Wu}\ \emph {et~al.}(2008)\citenamefont {Wu},
  \citenamefont {Artoni},\ and\ \citenamefont {Rocca}}]{wu.josab.25.1840.2008}%
  \BibitemOpen
  \bibfield  {author} {\bibinfo {author} {\bibfnamefont {J.-H.}\ \bibnamefont
  {Wu}}, \bibinfo {author} {\bibfnamefont {M.}~\bibnamefont {Artoni}}, \ and\
  \bibinfo {author} {\bibfnamefont {G.~C.~L.}\ \bibnamefont {Rocca}},\ }\href
  {\doibase 10.1364/JOSAB.25.001840} {\bibfield  {journal} {\bibinfo  {journal}
  {J. Opt. Soc. Am. B}\ }\textbf {\bibinfo {volume} {25}},\ \bibinfo {pages}
  {1840} (\bibinfo {year} {2008})}\BibitemShut {NoStop}%
\end{thebibliography}
%

\end{document}